\documentclass{article}
\usepackage[utf8]{inputenc}
\usepackage{amsmath}
\usepackage{comment}
\usepackage{graphicx}
\usepackage{amssymb}
\usepackage{xcolor}
\usepackage{soul}
\usepackage{epstopdf}
\usepackage{longtable}
\usepackage{appendix}
\usepackage{float}

\title{Modeling the Effect of Blunt Impact on Mitochondrial Dysfunction in Cartilage}
\author{Georgi I. Kapitanov, Bruce P. Ayati, James A. Martin }
\date{\today}

\begin{document}

\maketitle

\section{Introduction}\label{Intro}
Osteoarthritis (OA) is a degenerative disease characterized by thinning of the joint cartilage and is associated with disability and pain. A large portion of the cases are age-related -- progressive overloading of the cartilage, as well as the reduced abilities of the body to regenerate with age, result in cartilage lesions and joint inflammation \cite{Buckwalter:2005}. Osteoarthritis can also occur after a single forceful impact injury, in which case it is referred to as post-traumatic osteoarthritis (PTOA). We hypothesize that the biochemical processes associated with PTOA are the same as those that lead to OA, only occurring on a different time scale  while age-related OA can take decades to occur, PTOA can develop in a matter of months \cite{Anderson:2011, Buckwalter:2015}.\\

There are generally two hypotheses as to the biochemical source of degeneration in OA. One is the role of joint inflammation and particularly the disruption of the balance between pro- and anti- inflammatory cytokines in the joint \cite{Scanzello:2016}. Our group has been leading the modeling efforts in this area with several publications \cite{Graham:2012, Wang:2014, Wang:2015, Ayati:2016, Kapitanov:2016}. However, although non-invasive anti-inflammatory therapies reduce joint pain, they have not been found to slow the development of OA \cite{Kongtharvonskul:2016}.\\

Recent research efforts by the University of Iowa Department of Orthopedics and Rehabilitation have identified a different chondrocyte-centered hypothesis for the development of OA: oxidative stress and particularly the disruption of mitochondrial function as a result of joint overload \cite{Coleman:2015, Goetz:2016, Coleman2016}. The present work is the first attempt known to us to model this aspect of the underlying biochemistry after blunt impact. The article is organized as follows: Section \ref{model} presents the laboratory experimental set-up from already published work, the hypotheses, and the resulting mathematical model; Section \ref{results} presents our results; and Section \ref{disc} is a discussion.

\section{Model}\label{model}
This section details the experimental set-up we are modeling, the biological hypotheses involved, the resulting equations, and the computational work involved in solving and fitting the model to the experimental data.
\subsection{Laboratory Experiments}
Our modeling efforts revolve around laboratory experiments outlined in \cite{Martin:2009} and \cite{Coleman:2015}. Briefly, osteochondral explants (pieces of cartilage harvested from cattle), were subjected to high-energy blunt impacts of different magnitudes, comparable to those estimated to occur in serious joint injuries (7 J/cm$^2$ and 14 J/cm$^2$). The effects of impacts on cell viability within 72 h post-impact were recorded \cite{Martin:2009}. Effects of 7 J/cm$^2$ impacts on glucosaminoglycan (GAG) content, an indicator of cartilage stiffness and stability \cite{Coleman:2015}, were measured at 7 and 14 days post-impact. Metabolic activity as revealed by adenosine triphosphate (ATP) content was assessed at three different sites (impact, near impact, remote) at 24h and 48h post 7 J/cm$^2$ impact.

\subsection{Biological Hypothesis}
High energy impacts to cartilage cause local oxidative chondrocyte death \cite{Martin:2009}, as well
as a decline in ATP production (Figure 17.1). These effects were found to be related to excessive production of reactive oxygen species (ROS) by the mitochondrial electron transport chain, which causes damage to chondrocyte mitochondria and oxidative stress that inhibits glycolytic activity. The loss of ATP affects many cellular activities, but most notably it diminishes the production of GAGs, which undermines the stability of the cartilage matrix. Describing these processes mathematically and predicting their effects on the progression of PTOA is the subject of the present work.

\subsection{Mathematical Equations}
To understand the processes that occur at the impact area of the cartilage explant, we formulated a mathematical model to reflect the hypothesis outlined above. Since most of the data in \cite{Martin:2009} is relative (\% of viable cells, for example), we chose to use a unitless representation for these quantities. Generally, we consider 1 to be a level of the relevant variable that is considered optimal for cartilage function. Some deviations in that assumption are outlined later in this section for the choices of parameter values.\\
We include the following variables:
\begin{enumerate}
    \item $M(t)$: proportion of alive cells with functional (normal, undamaged) mitochondria in the cartilage explant impact area.
    \item $D(t)$: proportion of alive cells with dysfunctional (damaged, abnormal) mitochondria in the cartilage explant impact area. These cells are characterized by their release of double the amount of ROS as cells with functional mitochondria.
    \item $R(t)$: relative concentration of reactive oxygen species (ROS) in the cartilage explant impact area.
    \item $E(t)$: relative concentration of ATP (produced through glycolisis) in the cartilage explant impact area.
    \item $U(t)$: relative concentration of GAG (measure of cartilage stability) in the cartilage explant impact area.
\end{enumerate}
The following dynamical system describes the overall cellular/biochemical processes that occur after impact. Due to impact, we assume a sudden release of ROS (captured in different initial amount of ROS depending on the impact energy). This abnormal release of ROS in the explant causes oxidative stress. The effect of oxidative stress on the chondrocytes is transitioning of cells with normal, functional mitochondria into cells that have dysfunctional mitochondria; resulting in cell death or metabolic impairment. The difference between cells with functional mitochondria versus dysfunctional mitochondria is that the latter release twice the amount of ROS released by the former. The amount of ROS available within the cell determines the amount of ATP produced (explained in more detail later), and the amount of ATP available determines the production of GAG. We assume no cell proliferation -- none was observed during the experiments, and the experiment's time frame would suggest no significant number of new cells has been added.
\begin{align}\label{Orig_system}
\frac{dM}{dt} =& \underbrace{-k_SMS(R)}_{\text{mito damage due to ox. stress}}\\
\frac{dD}{dt} =& \underbrace{k_SMS(R)}_{\text{mito damage due to ox. stress}} - \underbrace{\delta_DDS(R))}_{\text{apoptosis due to ox. stress}}\\
\frac{dR}{dt} =& \underbrace{\alpha_M(M + k_DD)}_{\text{mito ROS release}} - \underbrace{\delta_{R}R}_{\text{ROS clearance}}\\
\frac{dE}{dt} =& \underbrace{f_E\left(\frac{R}{M + D + \epsilon}\right)}_{\text{ATP production}} - \underbrace{\delta_EE}_{\text{utilization}}\\
%\frac{dU}{dt} =& \underbrace{k_UE\frac{1}{1 - \lambda_UU}}_{\text{GAG through ATP}} - \underbrace{\delta_UU}_{\text{ECM catabolism}}
%\frac{dU}{dt} =& \underbrace{k_UE\frac{1}{1 + \lambda_UU}}_{\text{GAG through ATP}} - \underbrace{\delta_UU}_{\text{ECM catabolism}}
\frac{dU}{dt} =& \underbrace{k_UU(1 - \frac{1 + \lambda_U}{1 + \lambda_UE}U)}_{\text{GAG through ATP}}
\end{align}

The function $S(R)$ represents the effect of oxidative stress on the system. It only triggers when an excessive amount of ROS is present.
\begin{equation}
S(R) =
    \begin{cases}
 0 \text{ if } R \leq 1\\
 s_C(R-1)^{\alpha} \text{ if } R >1
\end{cases}
\end{equation}
The constant $s_C$ represents the direct effect of oxidative stress, represented by ROS being above the optimal level of 1, on the mitochondrial function and viability. We choose the constant $\alpha$ to be greater than 1. This ensures that $S(R)$ is smooth at $R = $1 and simplifies the equilibrium analysis.\\
The function $f_E(x)$ describes the energy (ATP) production. In our model $x$ is the ratio between available ROS and viable cells $R/(M + D + \epsilon)$. The parameter $\epsilon > 0$ is there to avoid division by 0.
\begin{equation}
f_E(x) =
    \begin{cases}
 0 \text{ if } x \leq 0 \text{ or } x \geq 2R_0\\
 \frac{k_E}{(x - R_0)^2 + \lambda_E} - \frac{k_E}{R_0^2 + \lambda_E} \text{ if } x\in(0, 2R_0)
\end{cases}
\end{equation}
What the function $f_E$ describes is that if the relative amount of ROS is below some optimal level $R_0$ or above it, then the energy production is lower than optimal. Furthermore, no available ROS ($R = 0$) or too much ROS ($R > 2R_0$) shuts down ATP production. This idea is presented in \cite{Coleman:2015}. A plot of the function can be seen in Figure \ref{fE_pic}.

\section{Results}\label{results}
This section includes mathematical analysis of the system equilibria and the computational results of the model.
\subsection{Mathematical Analysis}
In deterministic mathematical models, like the present one, the equivalent of statistical analysis done for experimental data or for stochastic models, is equilibrium analysis. We also analyze the local effect of parameter perturbations on the different variables through local sensitivity analysis. \\
Let $(M^*, D^*, R^*, E^*, U^*)$ denote an equilibrium solution to \eqref{Orig_system}. The only possible equilibrium solution occurs when $S(R^*) = 0$. Briefly, if $S(R^*) \neq 0$, $M^* = D^* = 0$, which implies that $R^* = 0$, but then $S(R^*)$ = 0, a contradiction. Therefore, $R^* \leq 1$. Furthermore, since no new cells are produced in the scope of this model and the explant assays, $M^* + D^* \leq 1$.\\
\subsubsection{Parameter relationships}\label{Par_rel_subs}
We assume that under homeostasis (undamaged cartilage), the values of cell with functional mitochondria, ROS, ATP, and GAG ($M, R, E$, and $U$ respectively) remains 1, while the value for cells with dysfunctional mitochondria, $D$, remains 0. The only reason for changes is disruption of this equilibrium due to an impact. To ensure this equilibrium, the following relationships between parameters were assumed
\begin{enumerate}
\item We assumed that in the function $f_E$ that $R_0 = 1/(1 + \epsilon)$ so as to produce the maximum amount of ATP when $R = 1$ and $M + D$ = 1.
%\item We considered a level of 90\% $M$ and 10\% $D$ to be optimal/normal for cartilage. This assumption requires that $\alpha_M = \delta_R/(0.9 + k_D0.1)$, since we seek an equilibrium  $R^* = 1$ when $M^* = 0.9$ and $D^* = 0.1$.
\item We considered a level of 100\% $M$ to be optimal/normal for cartilage. This assumption requires that $\alpha_M = \delta_R$, since we seek an equilibrium  $R^* = 1$ when $M^* = 1$.
\item {With the assumptions above, in order to produce an equilibrium $E^* = 1$ when $R^* = 1$, we assume that $\delta_E = \frac{k_E}{\lambda_E(1 + \lambda_E)}$.}
\end{enumerate}
%\item {In order to have an equilibrium $U^* = 1$ when $E^* = 1$, we assume that %$k_U = (1 - \lambda_U)\delta_U$.
%$k_U = (1 + \lambda_U)\delta_U$.}
The details of the equilibrium analysis are given in Appendix \ref{app_eq_analysis}. Briefly, the non-trivial equilibrium is stable and will be determined by the effect of the oxidative stress on the cell viability. In other words, we expect to reach a new homeostasis with lower levels of cell viability and appropriate levels of ROS, ATP, and GAG. No chaos is present in the system.

\subsection{ Numerical Results and Data Fitting}
System \eqref{Orig_system} was solved using the Matlab\texttrademark~function ode15s. The parameter values used for generating the results can be seen in Table \ref{Table_Parameter}. The data used for parametrization of our model can be seen in Table \ref{Data_table}. We note that the data is modified. In the experimental results in \cite{Martin:2009}, all explants had mean initial viability of 89\%, including control. If 89\% viability is normal for cartilage, we divided all the data by 89\% to get the normal viability to be 100\% (or 1 in the simulation calculations). In other words, the viability data was scaled.
%The initial conditions, in order $(M(0), D(0), R(0), E(0), U(0))$ for the no-impact simulation were (0.9, 0.1, 1, 1, 1).
The initial conditions, in order $(M(0), D(0), R(0), E(0), U(0))$ for the 7 J/cm$^2$ impact were (1, 0, 1.0202, 1, 1), and (1, 0, 1.036, 1, 1) for the 14 J/cm$^2$ impact simulation. We assumed that undamaged cartilage only contains cells with functional mitochondria, that ATP, and GAG content are optimal, and the impact increased the initial amount of ROS above 1, depending on the impact's energy. We fit all parameters, besides $R(0)$ for the 14 J/cm$^2$ impact using the 7 J/cm$^2$ data in Table \ref{Data_table} (cell viability, ATP, GAG). Then, using the parameters we found, we fit the initial amount of ROS after the 14 J/cm$^2$ impact to the cell viability data in that case. We used the Matlab particle swarm function for minimizing the error. The root mean square error for the fit to the 7 J/cm$^2$ data is 0.074, and to the 14 J/cm$^2$ data is 0.123. The results are presented in Figures \ref{Mito_fit_7} to \ref{GAG_data}. The total cell viability (functional plus dysfunctional mitochondria) fits well with the cellular viability presented in \cite{Martin:2009}, as evident from Figures \ref{Mito_fit_7} and \ref{Mito_fit_14}. The ATP simulation also fits well with the available data from \cite{Coleman:2015} as seen in Figure \ref{ATP_data}. The GAG simulation also fit well with the given data (Figure \ref{GAG_data}). Overall, the model seems to capture the biochemical dynamics of the impact site of the cartilage explant.\\

 \begin{longtable}[ht]{|c|c|}
         \hline
         Parameter & Value\\
         \hline
         $\kappa_S$ & 2.7938\\
         $\delta_D$ & 9.9626\\
         $\delta_R$ & 0.0727\\
         $\kappa_E$ & 0.0961\\
         $\lambda_E$ & 0.0418\\
         $k_U$ & 5.0\\
         $\lambda_U$ & 0.3387\\
         $s_C$ & 9.517\\
         $R_0$ & 1/(1 + $\epsilon$)\\
         $\epsilon$ & 10$^{-4}$\\
         $\alpha$ & 1 + $\epsilon$\\
         \hline
    \caption{ Table of parameters}
\label{Table_Parameter}
 \end{longtable}

 \begin{longtable}[h!]{|c c c|c c|c c|}
         \hline
        \multicolumn{3}{|c|}{Cell viability, \%}&\multicolumn{2}{|c|}{GAG, \% non-impact}&\multicolumn{2}{|c|}{ATP, \% Control average}\\
         time, h& 7 J/cm$^2$ & 14 J/cm$^2$ & time, d& 7 J/cm$^2$ & time, h & 7 J/cm$^2$\\
         0 & 100$\pm$ 8 & 100$\pm$ 8 & 7  & 81$\pm$ 4& 24      & 18$\pm$ 18\\
         1 & 80$\pm$ 9 & 73 $\pm$ 4 & 14 & 87$\pm$ 10& 48      & 30$\pm$ 19\\
         2 & 74$\pm$ 7 & 71$\pm$ 8 &&&&\\
         4 & 60 $\pm$ 13 & 60$\pm$ 2 &&&&\\
         6 & 65$\pm$ 6   & 43 $\pm$ 2 &&&&\\
         12& 51$\pm$ 7   & 42$\pm$ 7 &&&&\\
         24& 52$\pm$ 7   & 39$\pm$ 4 &&&&\\
         48& 47$\pm$ 7   & 39$\pm$ 7 &&&&\\
         72& 52$\pm$ 7   & 44$\pm$ 7 &&&&\\
         \hline
        \caption{ Data used in parameter estimation. Standard deviation is given after the mean as $\pm$.}
\label{Data_table}
 \end{longtable}
%The error from the 7J impact was 0.0651 and from the 14J impact was 0.1206.

\subsection{Sensitivity Analysis}
Let us denote by $S_{\text{par},\text{var}}$ the effect of the parameter \emph{par} on the variable \emph{var}. Standard methods of local sensitivity analysis boil down to solving a set of differential equations with respect to $S_{\text{par},\text{var}}$, namely
\[\frac{d\vec{S}_{\text{par}}}{dt} = J\cdot \vec{S}_{\text{par}} + F, \]
where $\vec{S}_{\text{par}}$ is the vector of $S_{\text{par},\text{var}}$ with respect to each variable, $J$ is the Jacobian matrix, and $F$ is a vector of partial derivatives of the corresponding variable with respect to the parameter of interest.\\
The parameters we want to analyse are $k_S, \delta_D, \delta_R, s_C, k_E, \lambda_E, k_U$, and  $\lambda_U$, as well as the initial conditions for each variable, $M(0), D(0), R(0), E(0), U(0)$. The method is outlined in \cite{Atherton:1975}.
The relative local effect was measured by $S_{\text{par,var}}(t)/\text{var}(t)$. None of the parameters affect any of the variables locally. The initial conditions had some effect, although none of them had a local effect on $U$. $E(0)$ and $U(0)$ did not affect any of the variables. The effect of changes in $M(0)$ on $M(t)$ is constantly 1, on $D$ is 0, and on $R$ and $E$ can be seen in Figures \ref{sens_R_M} and \ref{sens_E_M}. $D(0)$ does not affect $M$ and $D$ and its effect on $R$ and $E$ can be seen in Figures \ref{sens_R_D} and \ref{sens_E_D}. Changes in $R(0)$ affects $R$ and $E$, as seen in Figures \ref{sens_R_R} and \ref{sens_E_R}, respectively.\\

\section{Discussion}\label{disc}
We constructed a model of the effects of oxidative stress on the energy production and proteoglycan release of a cartilage explant after a blunt impact. The model considered the effect of the impact on the mitochondrial ROS release and the resulting disruption in ATP production, which in turn negatively affects GAG release and cartilage structure.\\
The model's results fit well with the results of the laboratory experiments both qualitatively and quantitatively. The simulations seem to capture well the cell viability dynamics, the amount of ATP available at the impact site post injury, as well as the GAG content. Particularly, they capture the recovery in ATP production after the initial cell death and disruption. They capture, to an extent, the GAG production recovery as well.\\

The fact that the model's outcomes are not sensitive to the local perturbations of the variables implies that our system is stable, as suggested by the equilibrium analysis. While a lot of biological systems exhibit chaos, we expect certain outcomes in this phenomenon, in particular for a significant impact on a cartilage explant to reduce the cellular viability and disrupt the production of ATP. Our system is sensitive to the initial conditions, which is understandable. The amount of ROS gradually becomes independent of the initial burst of ROS due to the impact, and dependent on the current viability of both cell types. This effect is seen in Figures \ref{sens_R_M}, \ref{sens_R_D}, and \ref{sens_R_R}. At the same time, the amount of initial ROS has a significant effect on the amount of cells with dysfunctional mitochondria, $D$, which is not surprising given that we assume that oxidative stress leads to apoptosis. Significantly less of an effect was observed in the sensitivity of cells with functional mitochondria to the initial amount of ROS, so it was assumed to be 0.\\

Several limitations of our modeling efforts should be addressed. Scarcity of longitudinal data for ATP production and GAG availability means the model results only vaguely support the underlying hypothesis about the effect of the post-impact oxidative stress on the biochemical functions in articular cartilage, but they do not allow us to conclude that the dynamics we describe are entirely accurate. Furthermore, the whole model is non-dimensional and the estimated parameters non-mechanistic, which makes the model only appropriate for estimating relative levels of the biochemical compounds. More data and measurements would be needed for addressing these issues.\\

The parameters themselves were estimated and the error found may be a local minimum, rather than a global one, so other parameter sets might give us a similar fit. However, the idea that the impact changes the initial amount of ROS released in the tissue seems to work to validate the viability of cells after the 14 J/cm$^2$ impact. More data at that stronger impact level would be needed to validate our ATP and GAG predictions. Our predictions for the amount of ROS present, both in the explant impact area, and per cell in the impact area seem to qualitatively capture expectations, namely high amounts of initial ROS, which level off eventually, as seen in \cite{Goodwin:2010}.\\

A major result presented in \cite{Martin:2009} suggests the positive effect of antioxidants, N-acetylcysteine (NAC) particularly, on the post-impact cellular viability and overall cartilage stability, as measured by GAG content. The fact that treating the cartilage explant with NAC results in mitigating the effects of the blunt impact, leads to the conclusion that reducing oxidative stress and mitochondrial dysfunction post-impact is a viable option for preventing the development of OA. Modeling the effect of NAC and the timing of its application will be the subject of further work.\\

Work on creating and implementing \emph{in silico} models like the one presented here may have a significant role in predicting the harmful effect of impact on cartilage explants and eventually translate to predicting post-impact patient outcomes. Using mathematical models to describe and quantify the biochemical reactions that lead to cartilage damage after an impact, may eventually remove the need to run a large portion of laboratory experiments. An ODE system such as \eqref{Orig_system} is easily encapsulated in code (e.g. Matlab) that can be used directly in the lab to predict experimental outcomes. The computations of the solutions the system take on the order of seconds on contemporary desktop equipment, which can save significant experimental time and resources relative to conducting expensive, time-consuming, and error-prone experiments. However, in the near term more experiments are needed to inform models and for creating an accurate map of the important biochemical interactions.

\bibliographystyle{siam}
\bibliography{Blunt_mito_bib}

\newpage
\section*{Figures}

 %% Figure %%
\begin{figure}[H]
\centering
\includegraphics[width=0.8\textwidth]{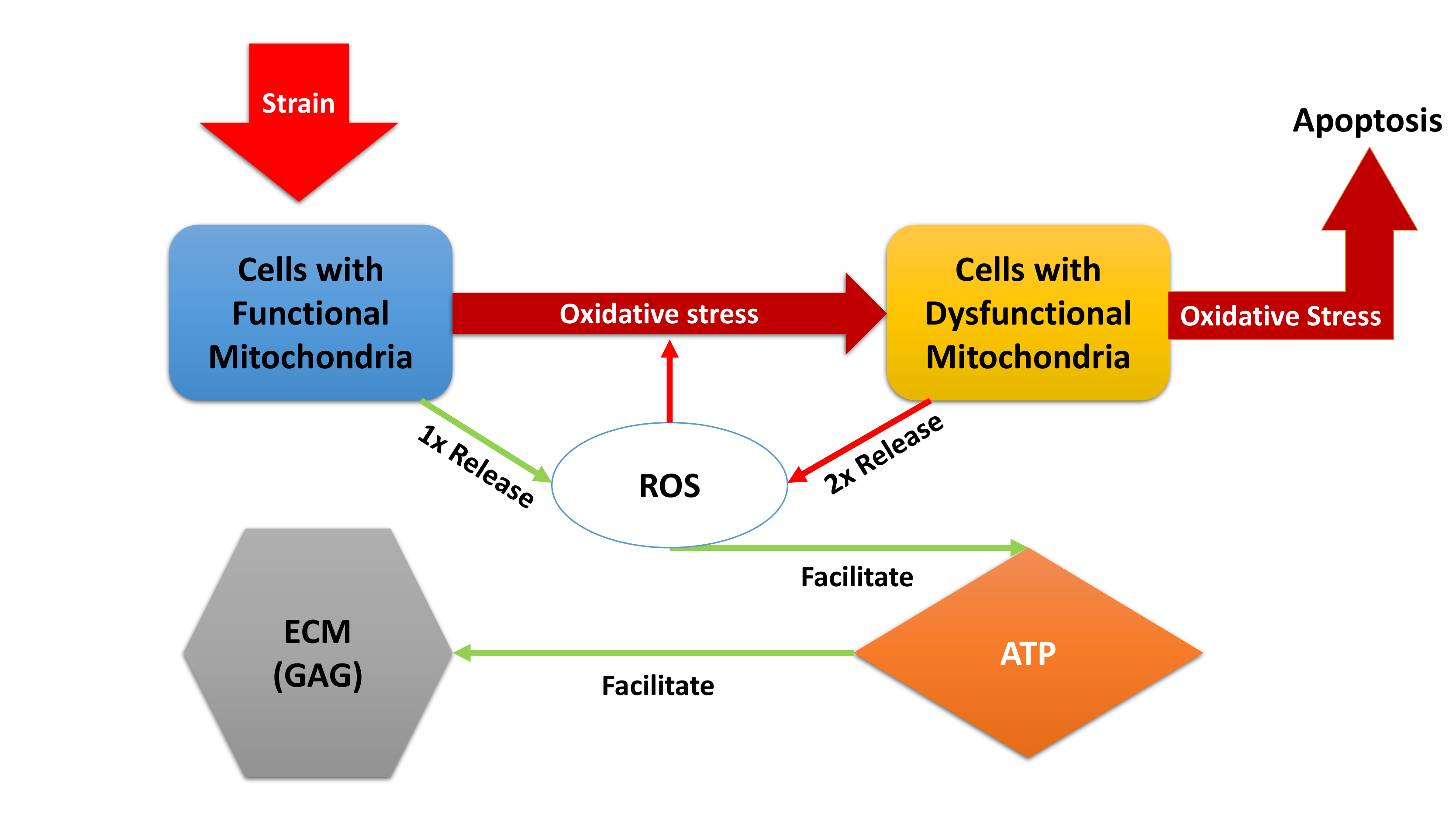}
\caption{A diagram of the dynamics expressed in \ref{Orig_system}. External strain from the blunt impact causes cells with functional mitochondria to transition into cells with dysfunctional mitochondria and cells with dysfunctional mitochondria to go into apoptosis. Cells with dysfunctional mitochondria release twice the amount of reactive oxygen species (ROS) as normal cells, which further affects the oxidative stress. ROS is used in production of ATP, which in turn is utilized for the release of glycosaminoglycans (GAG), which strengthen the ECM. }\label{Diagram}
\end{figure}
%% Figure end %%

%% Figure %%
\begin{figure}[H]
\centering
\includegraphics[width=0.8\textwidth]{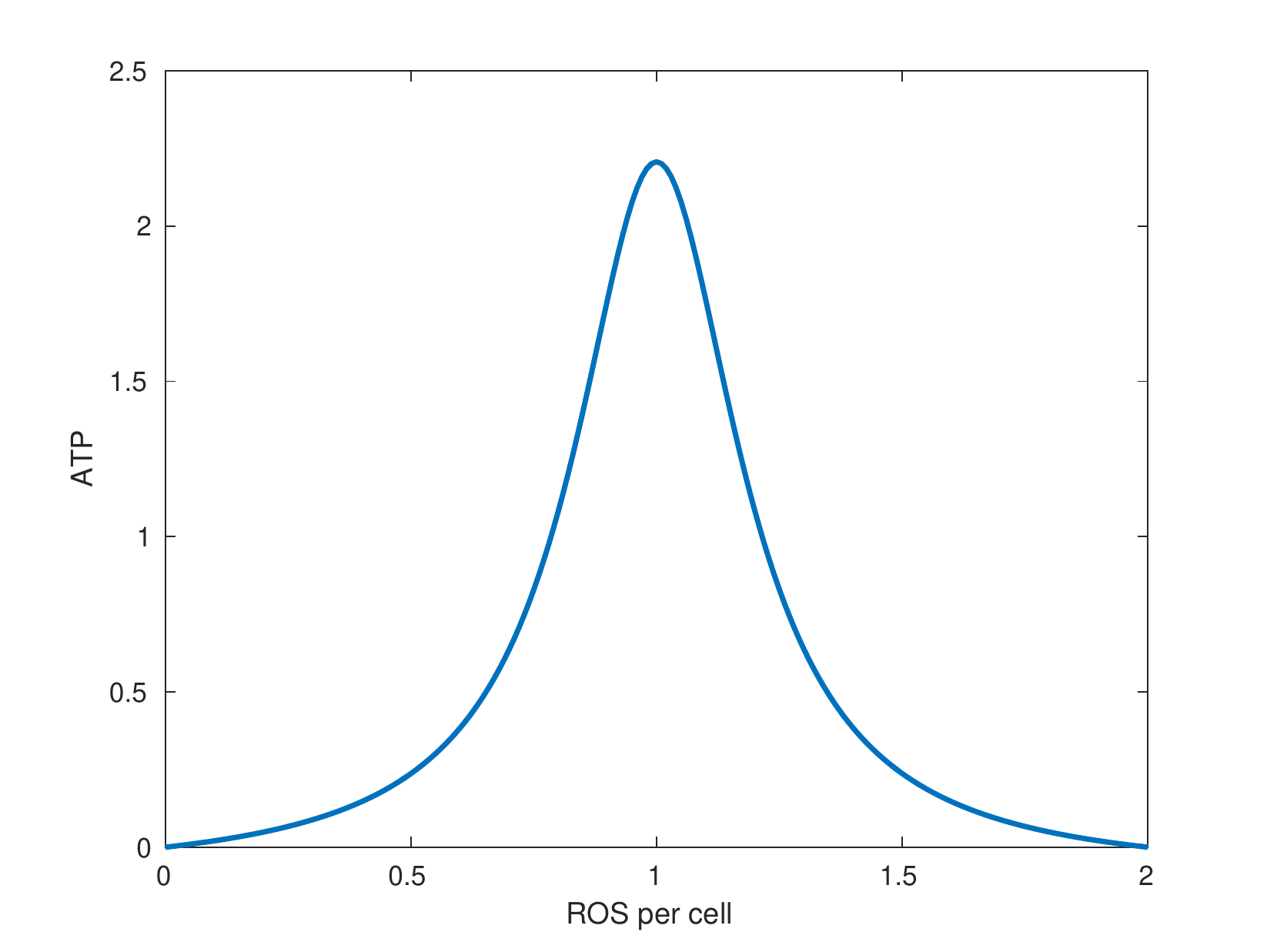}
\caption{The form of the function $f_E$. The peak (optimal) ATP production occurs when the amount of ROS per cell ($R/(M+D)$) is equal to $R_0$. In the picture above $R_0 = 1/(1.0001)$}\label{fE_pic}
\end{figure}
%% Figure end %%

 %% Figure %%
\begin{figure}[H]
\centering
\includegraphics[width=0.8\textwidth]{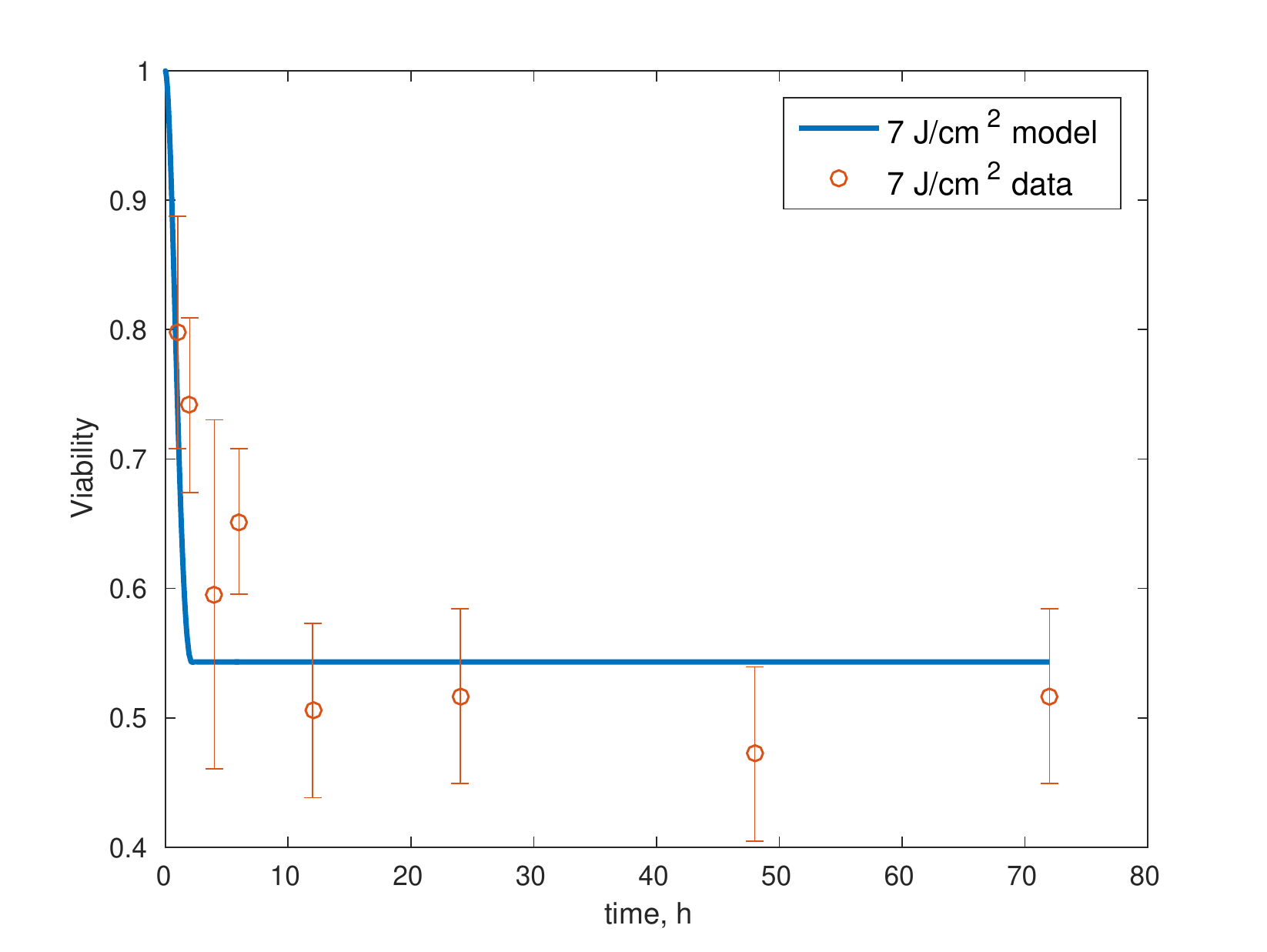}
\caption{Relative proportion of live cells after an impact of 7 J/cm$^2$ and its fit to available cell viability data from \cite{Martin:2009} (open circles).}\label{Mito_fit_7}
\end{figure}
%% Figure end %%

 %% Figure %%
\begin{figure}[H]
\centering
\includegraphics[width=0.8\textwidth]{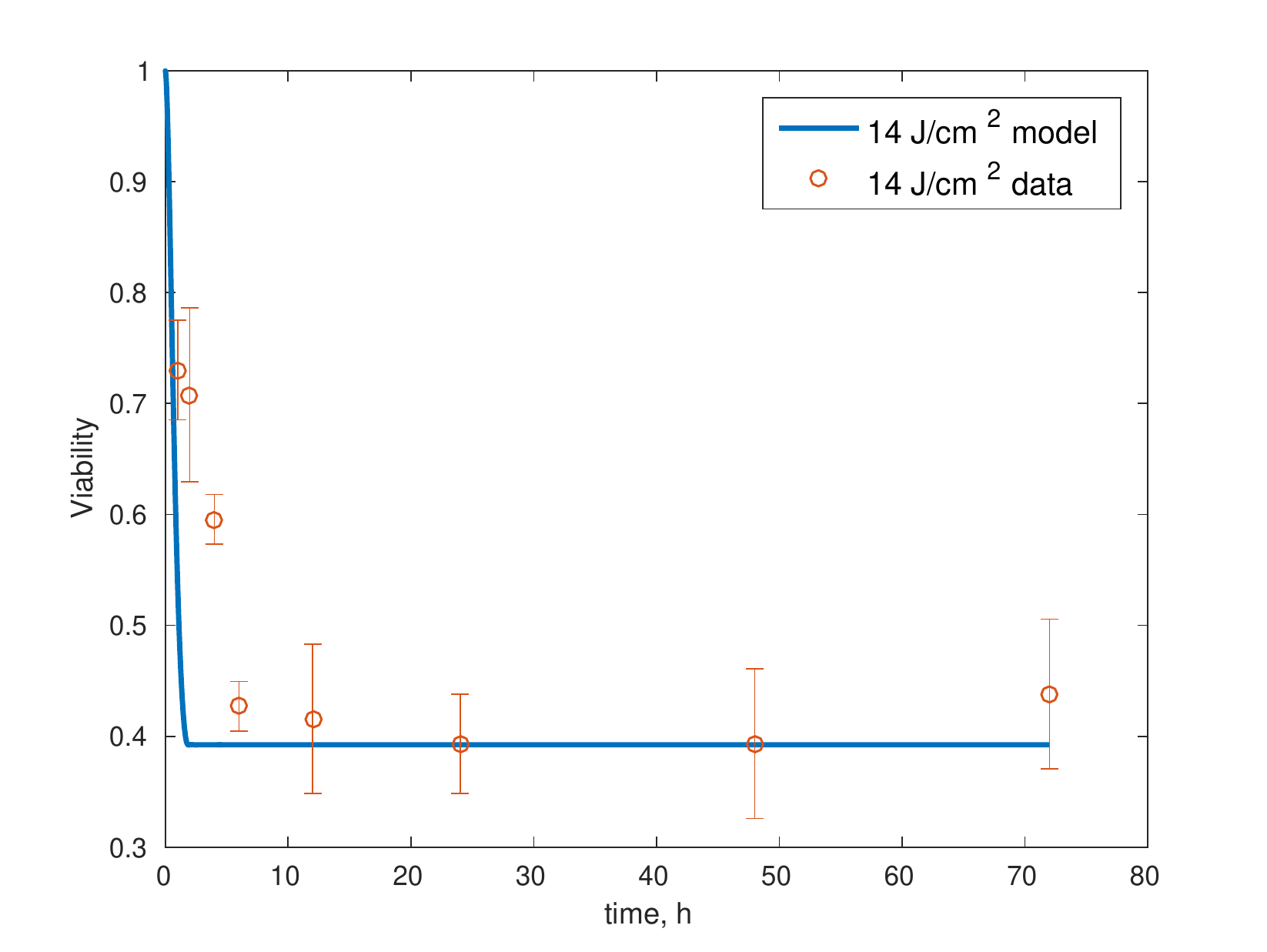}
\caption{Relative proportion of live cells after an impact of 14 J/cm$^2$ and its fit to available cell viability data from \cite{Martin:2009} (open circles).}\label{Mito_fit_14}  \end{figure}
%% Figure end %%

 %% Figure %%
\begin{figure}[H]
\centering
\includegraphics[width=0.8\textwidth]{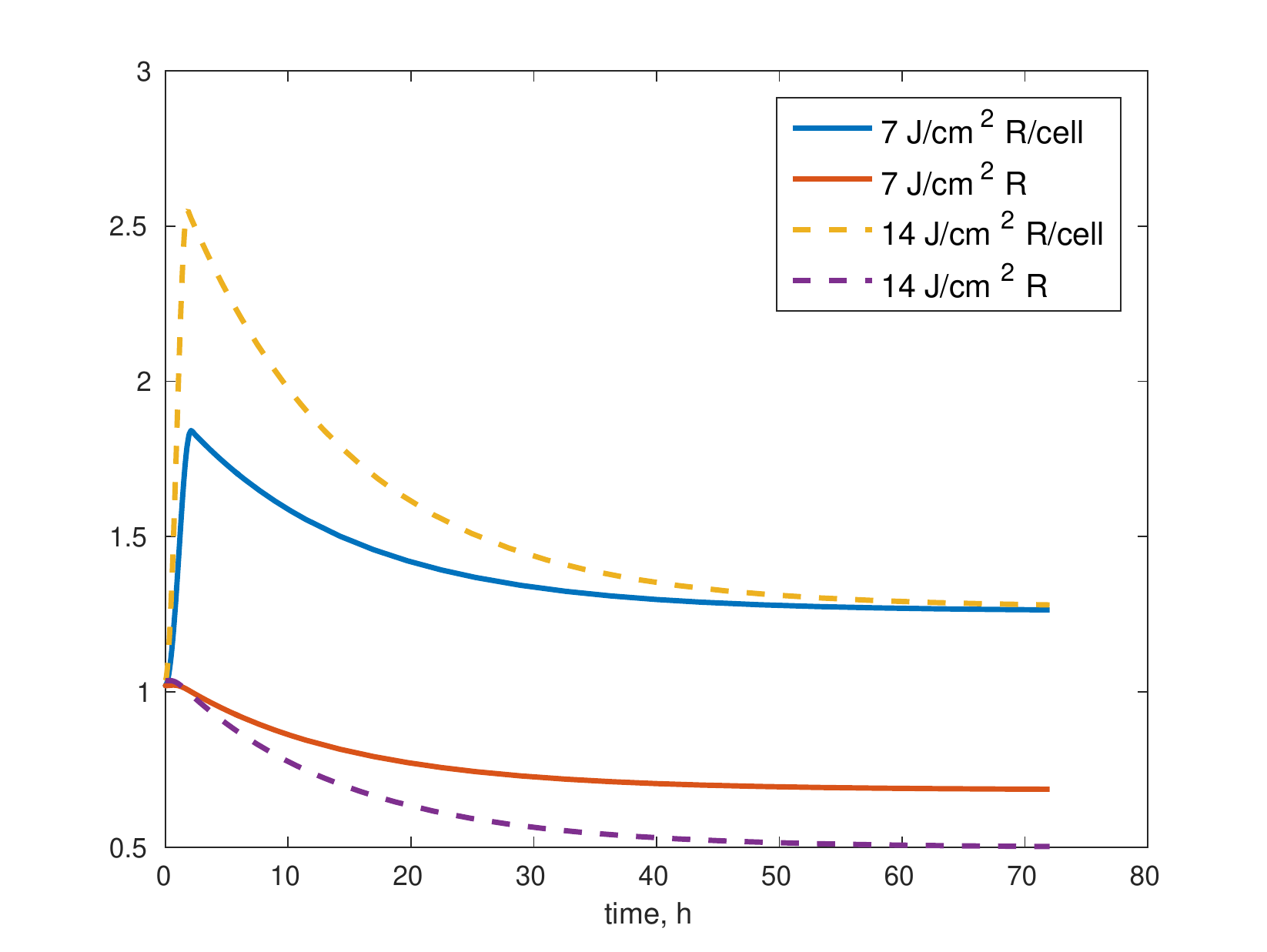}
\caption{Projected relative concentration of available ROS after 7 J/cm$^2$ impact,  and 14 J/cm$^2$ impact, after 72h.}\label{ROS_70h}  \end{figure}
%% Figure end %%

 %% Figure %%
\begin{figure}[H]
\centering
\includegraphics[width=0.8\textwidth]{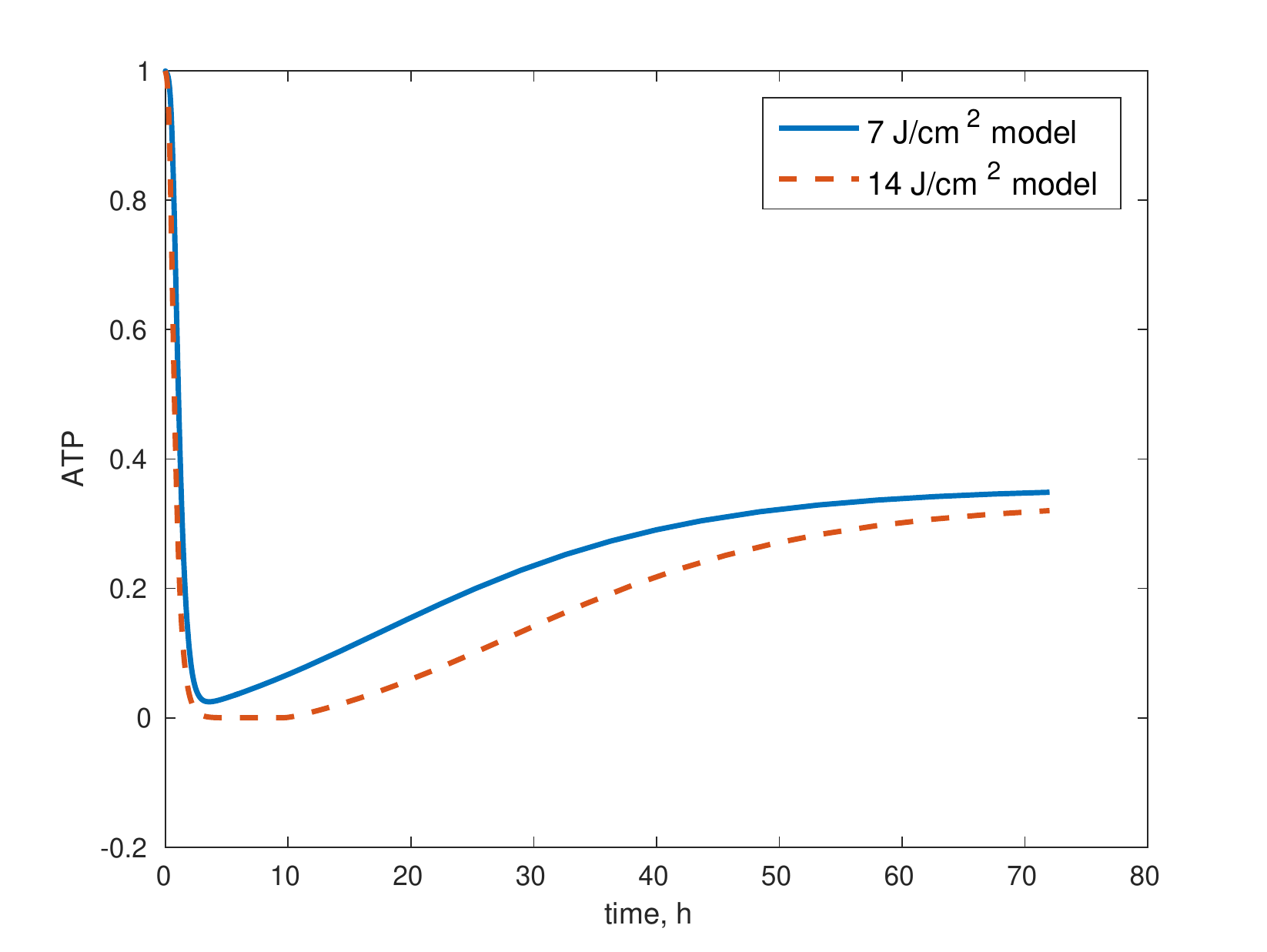}
\caption{Projected relative concentration of available ATP after 7 J/cm$^2$ impact,  and 14 J/cm$^2$ impact, after 72h.}\label{ATP_70h}  \end{figure}
%% Figure end %%

%% Figure %%
\begin{figure}[H]
\centering
\includegraphics[width=0.8\textwidth]{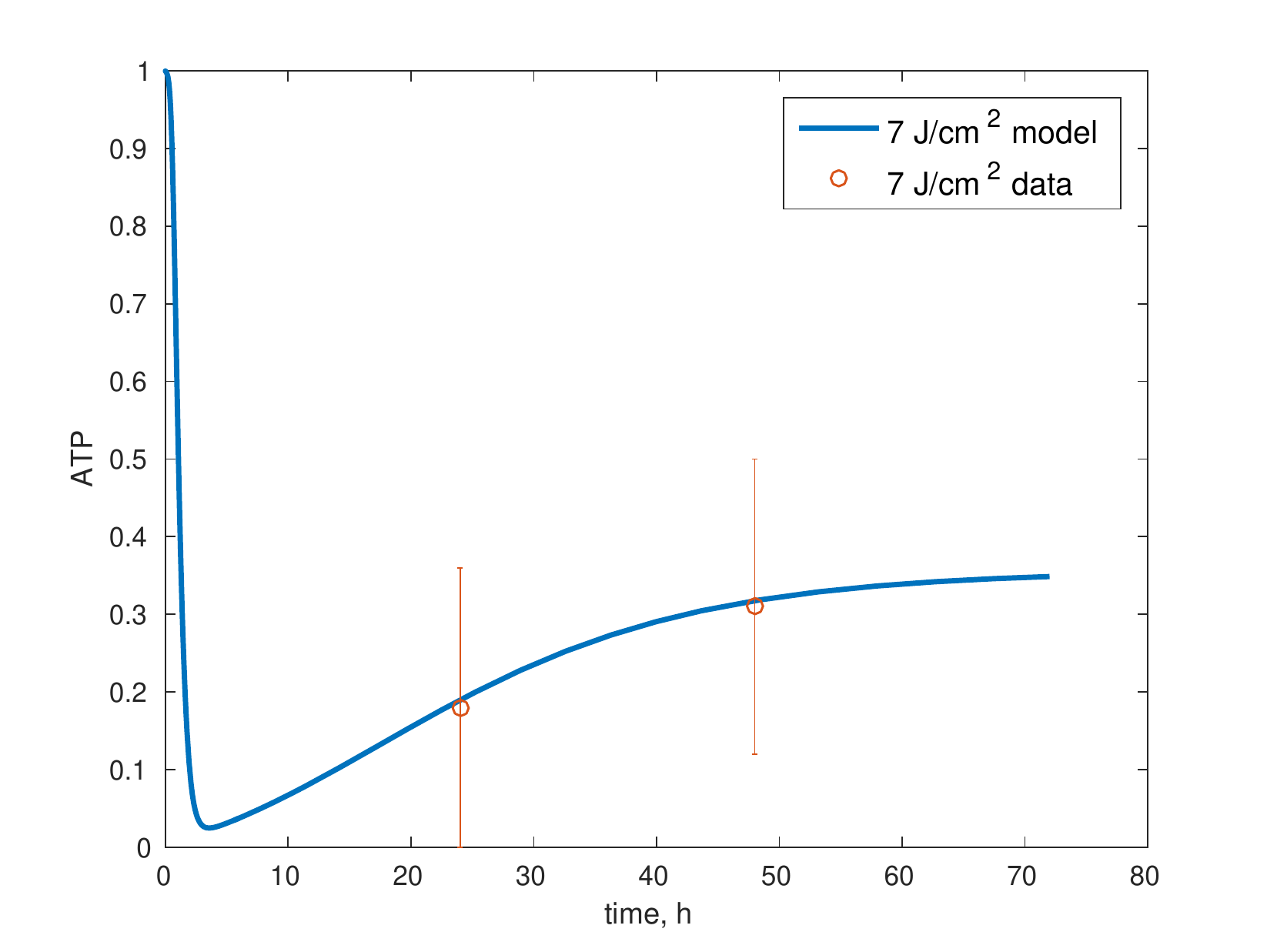}
\caption{Relative concentration of ATP after an impact of 7 J/cm$^2$ and its fit to available ATP data from \cite{Coleman:2015}.}\label{ATP_data}  \end{figure}
%% Figure end %%

%% Figure %%
\begin{figure}[H]
\centering
\includegraphics[width=0.8\textwidth]{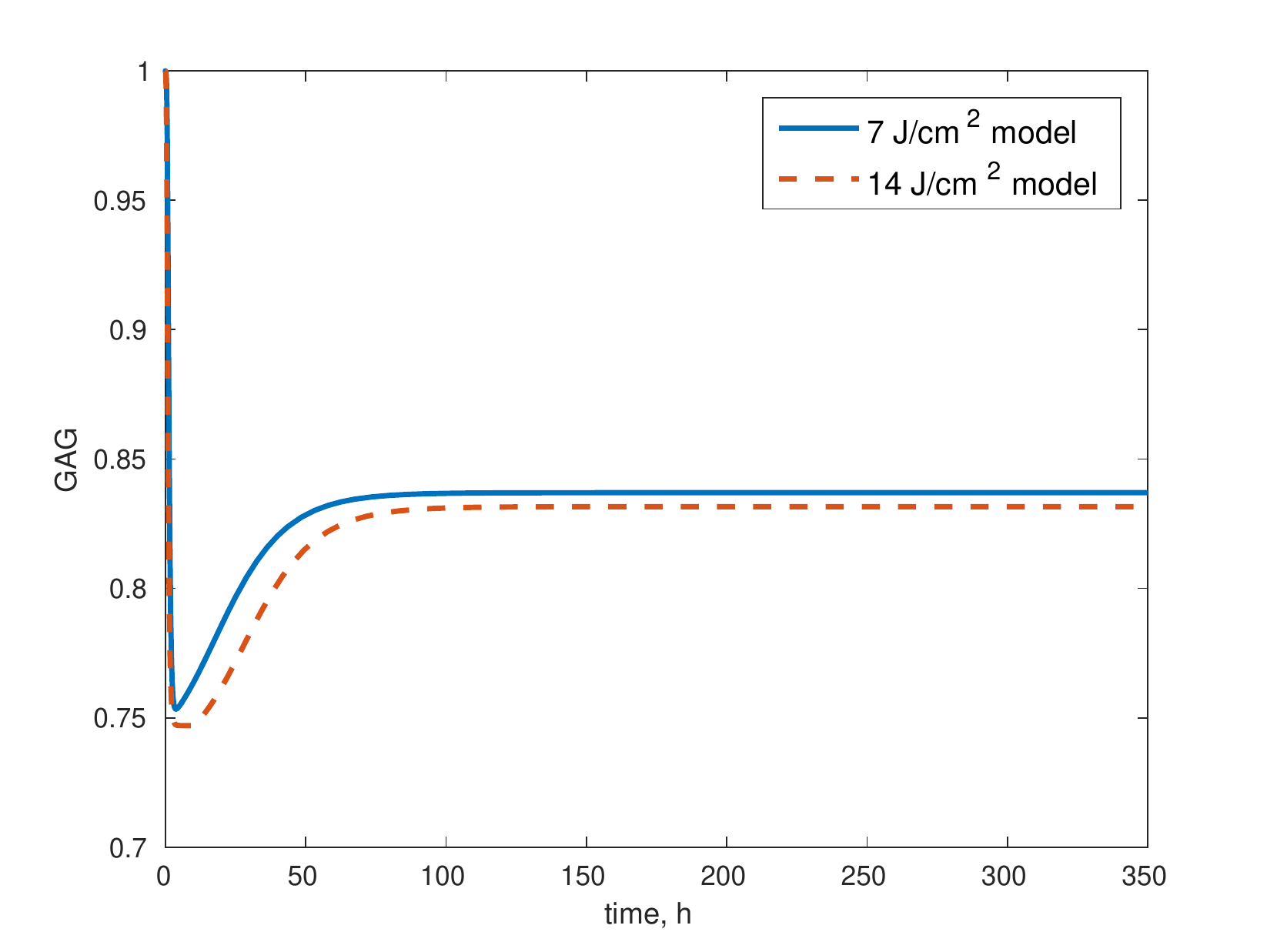}
\caption{Projected relative concentration of GAG after 7 and 14 J/cm$^2$ impact, after 250h.}\label{GAG_2w}  \end{figure}
%% Figure end %%

%% Figure %%
\begin{figure}[H]
\centering
\includegraphics[width=0.8\textwidth]{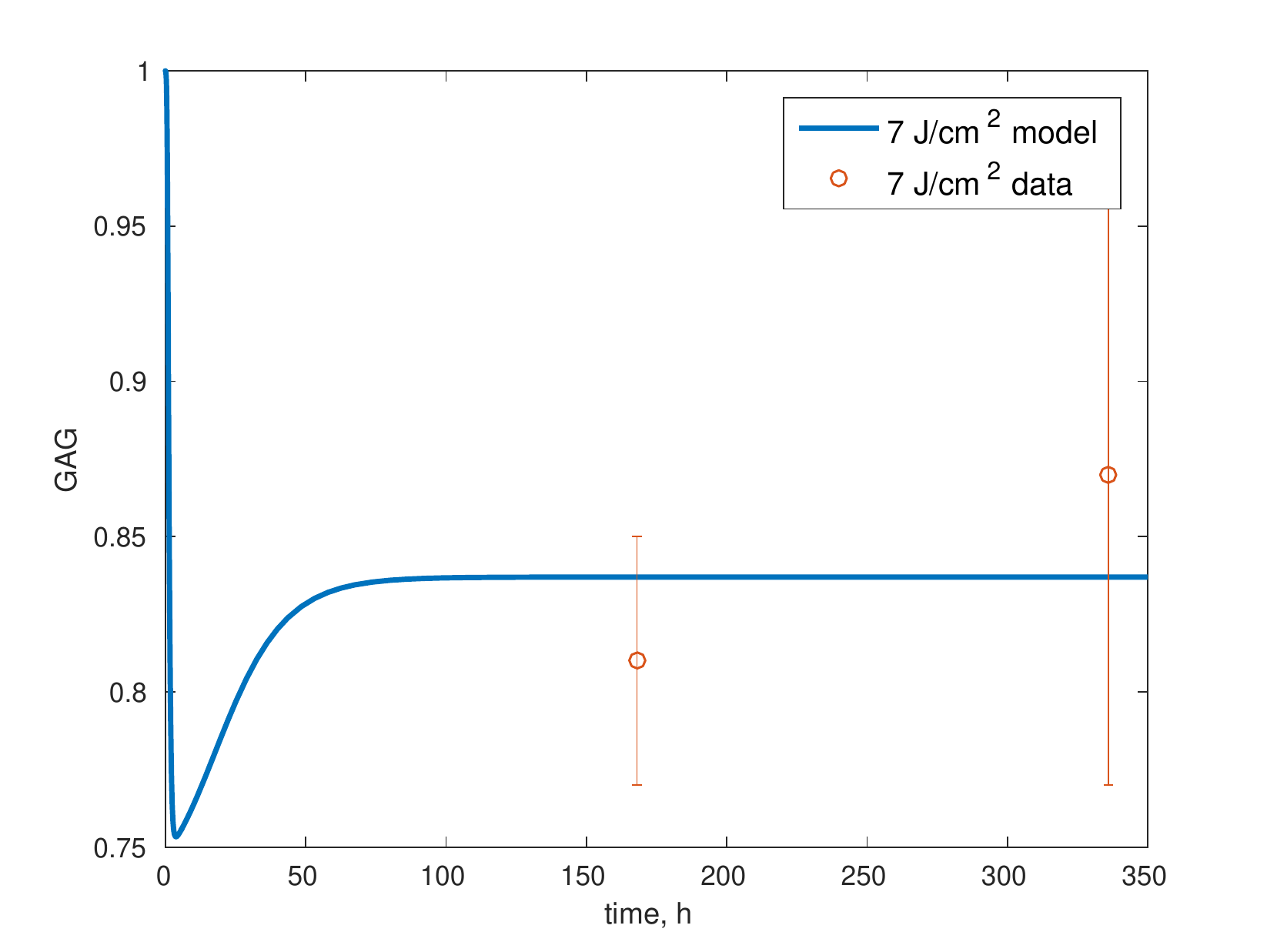}
\caption{Relative concentration of GAG after an impact of 7 J/cm$^2$ and its fit to available GAG data from \cite{Martin:2009}.}\label{GAG_data}  \end{figure}
%% Figure end %%

%% Figure %%
\begin{figure}[H]
\centering
\includegraphics[width=0.8\textwidth]{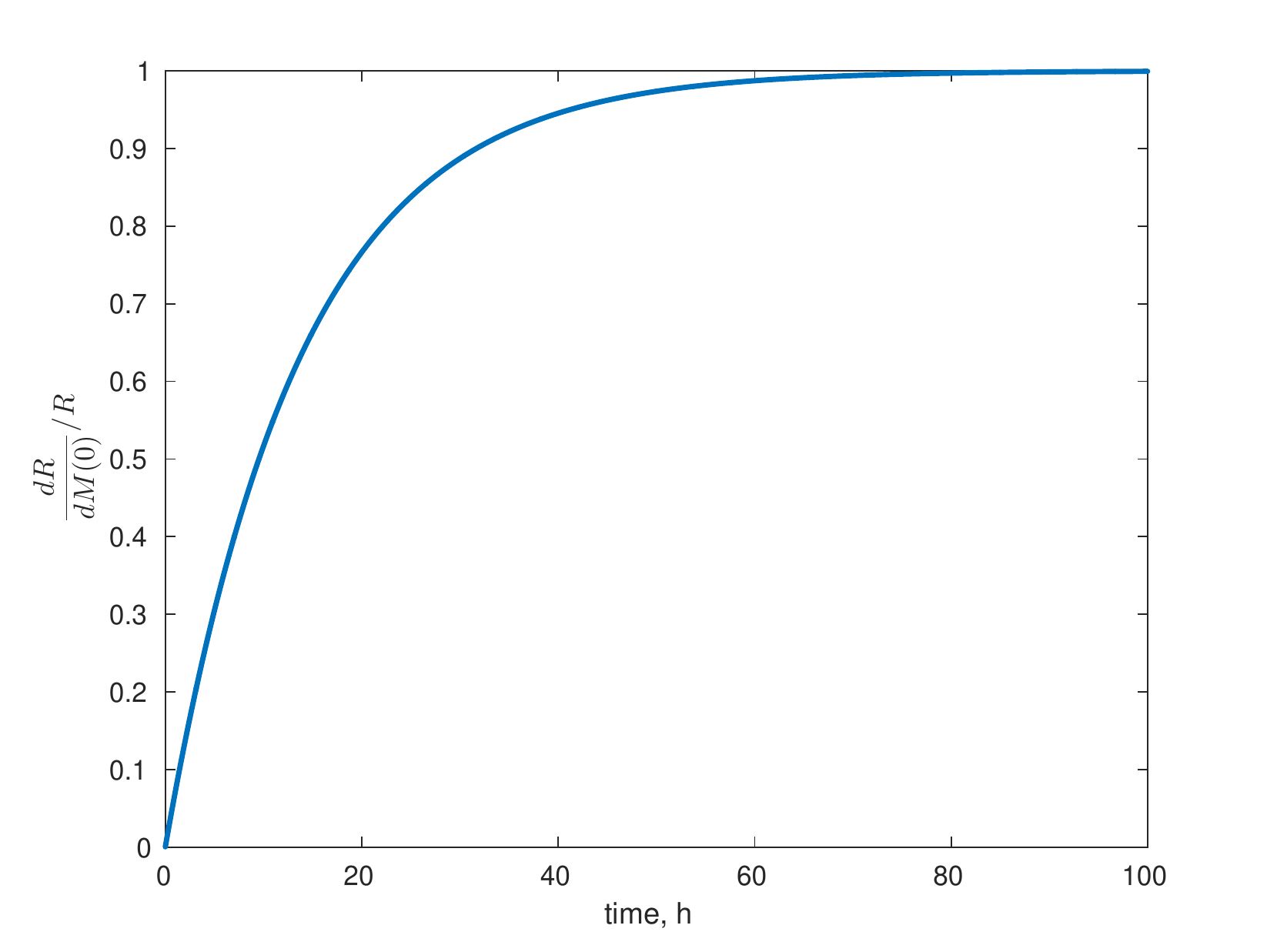}
\caption{Relative sensitivity of ROS to the initial proportion of cells with functional mitochondria, $M(0)$.}\label{sens_R_M}  \end{figure}
%% Figure end %%

%% Figure %%
\begin{figure}[H]
\centering
\includegraphics[width=0.8\textwidth]{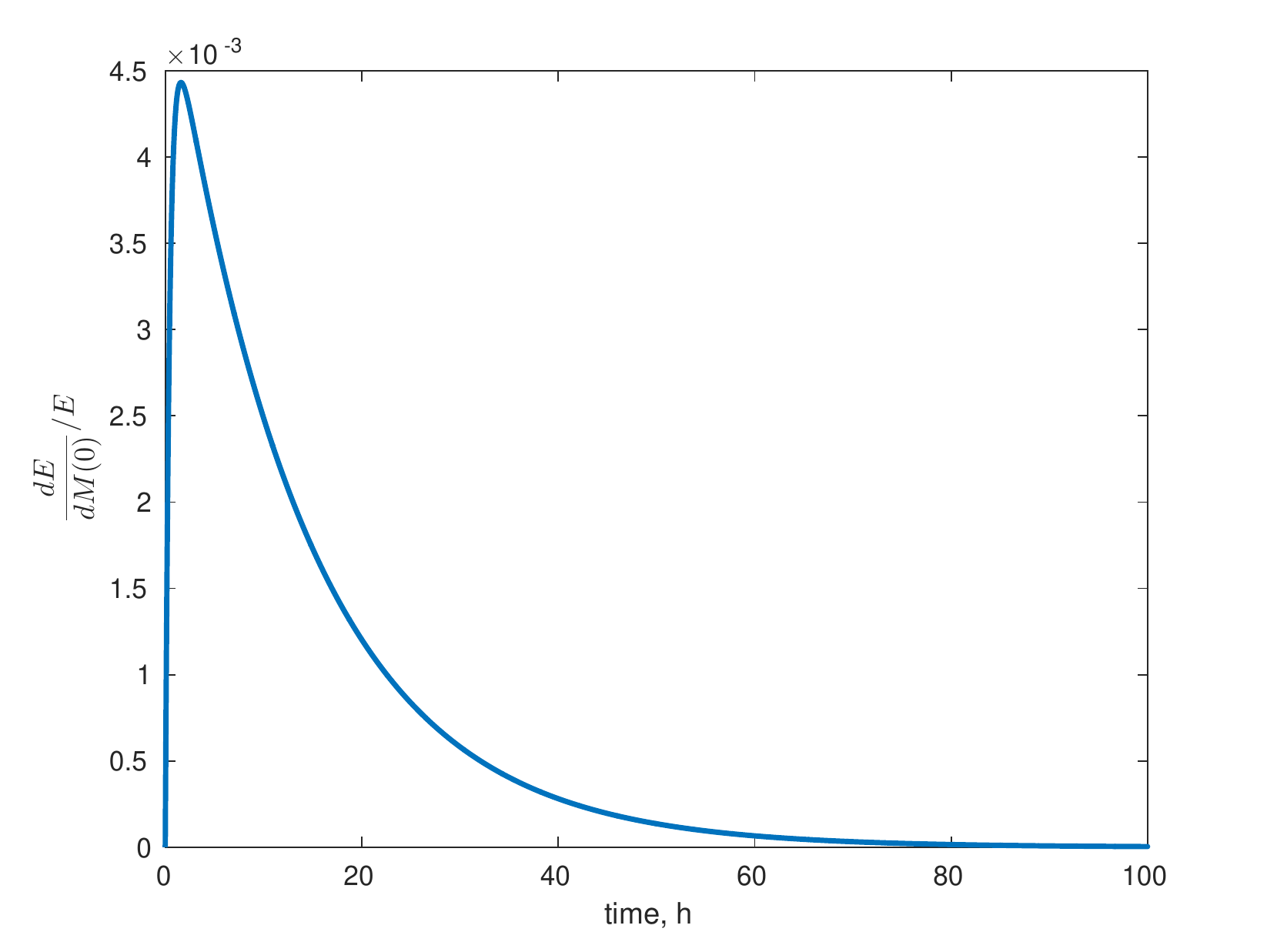}
\caption{Relative sensitivity of ATP to the initial proportion of cells with functional mitochondria, $M(0)$.}\label{sens_E_M}  \end{figure}
%% Figure end %%

%% Figure %%
\begin{figure}[H]
\centering
\includegraphics[width=0.8\textwidth]{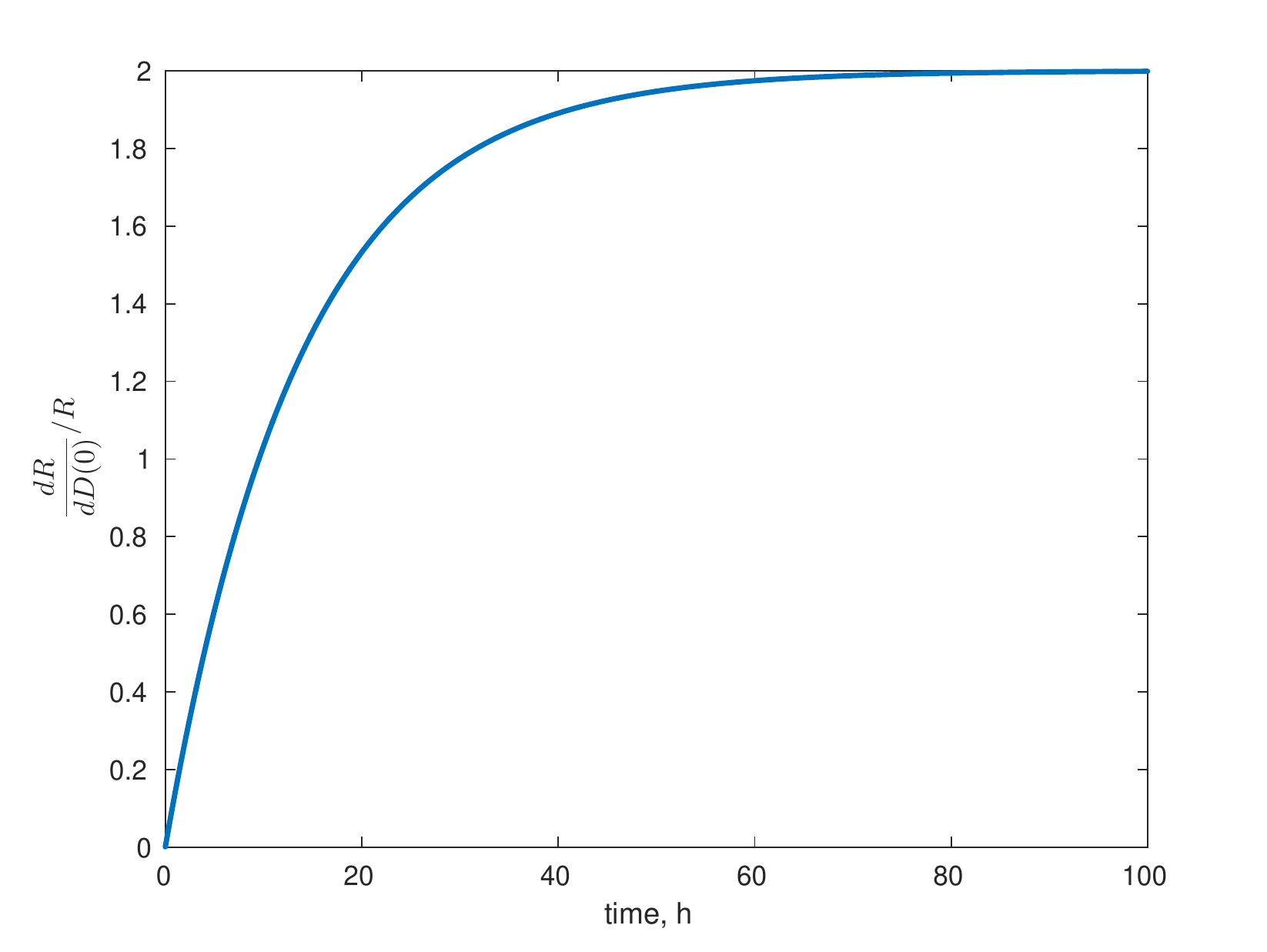}
\caption{Relative sensitivity of ROS to the initial proportion of cells with dysfunctional mitochondria, $D(0)$.}\label{sens_R_D}  \end{figure}
%% Figure end %%

%% Figure %%
\begin{figure}[H]
\centering
\includegraphics[width=0.8\textwidth]{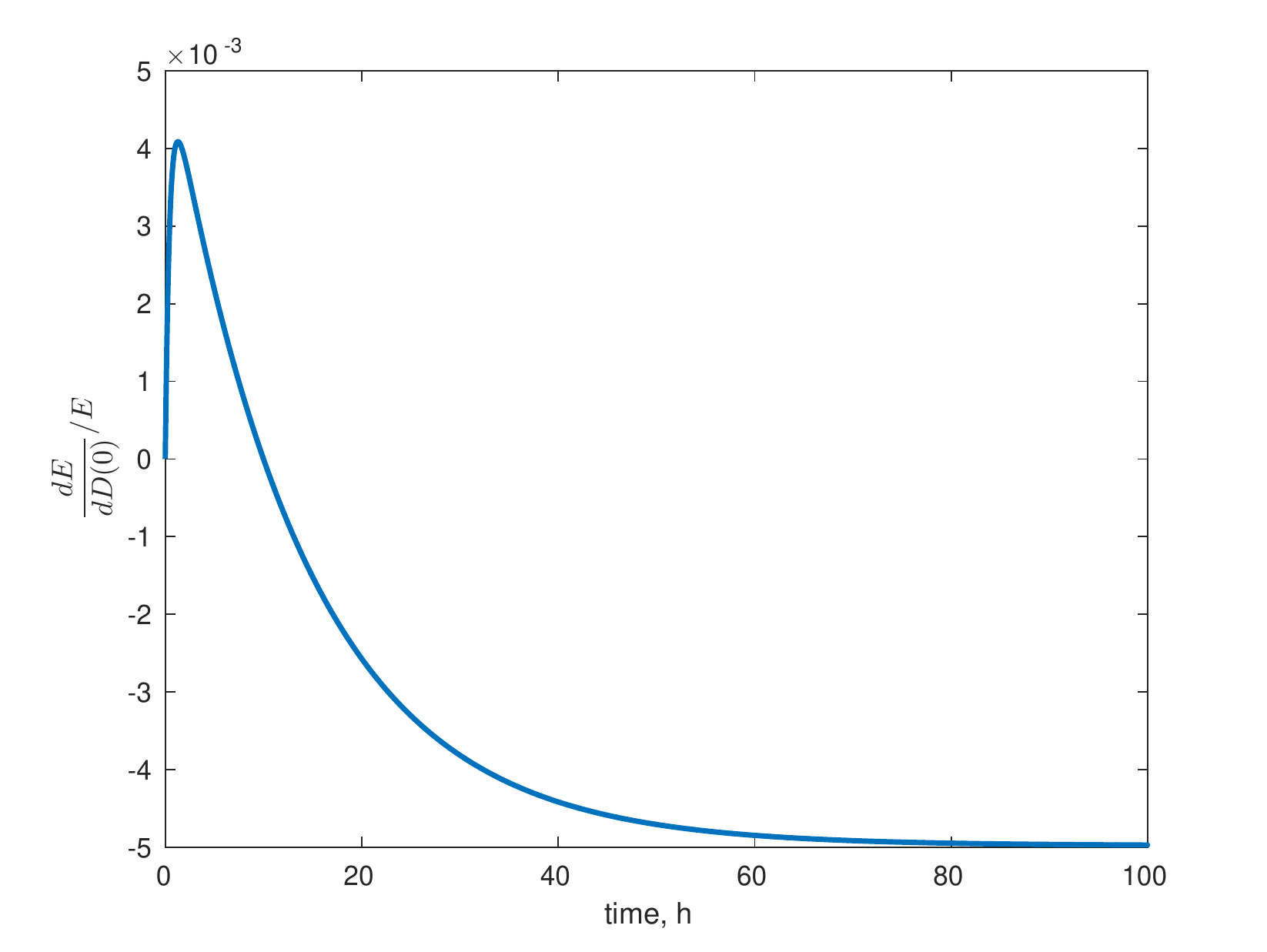}
\caption{Relative sensitivity of ATP to the initial proportion of cells with dysfunctional mitochondria, $D(0)$.}\label{sens_E_D}  \end{figure}
%% Figure end %%

%% Figure %%
%\begin{figure}[H]
%\centering
%\includegraphics[width=0.8\textwidth]{Rel_sens_D_to_R.eps}
%\caption{Relative sensitivity of cells with dysfunctional mitochondria to the initial concentration of ROS, $R(0)$.}\label{sens_D_R}  \end{figure}
%% Figure end %%

%% Figure %%
\begin{figure}[H]
\centering
\includegraphics[width=0.8\textwidth]{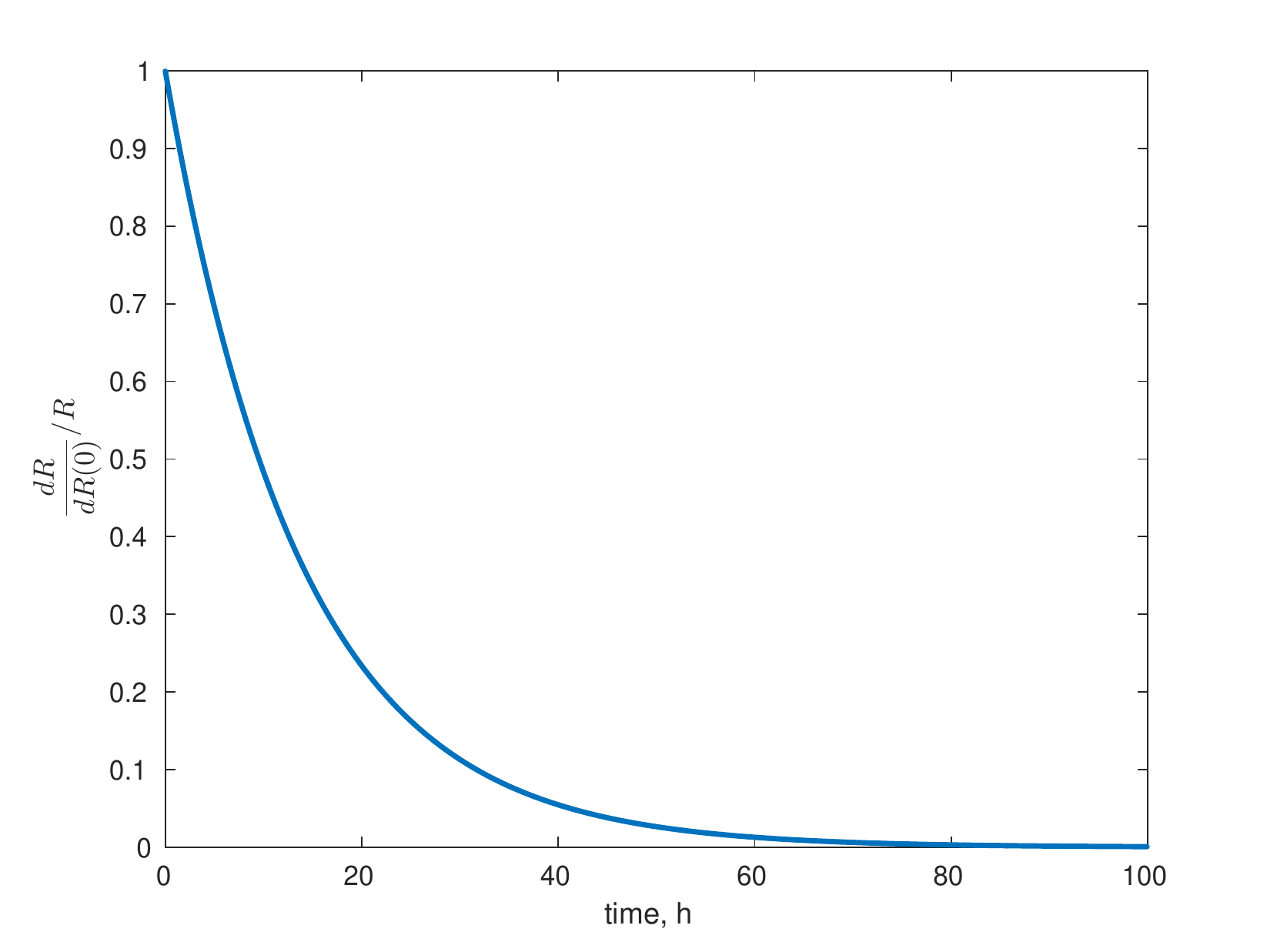}
\caption{Relative sensitivity of ROS to the initial concentration of ROS, $R(0)$.}\label{sens_R_R}  \end{figure}
%% Figure end %%

%% Figure %%
\begin{figure}[H]
\centering
\includegraphics[width=0.8\textwidth]{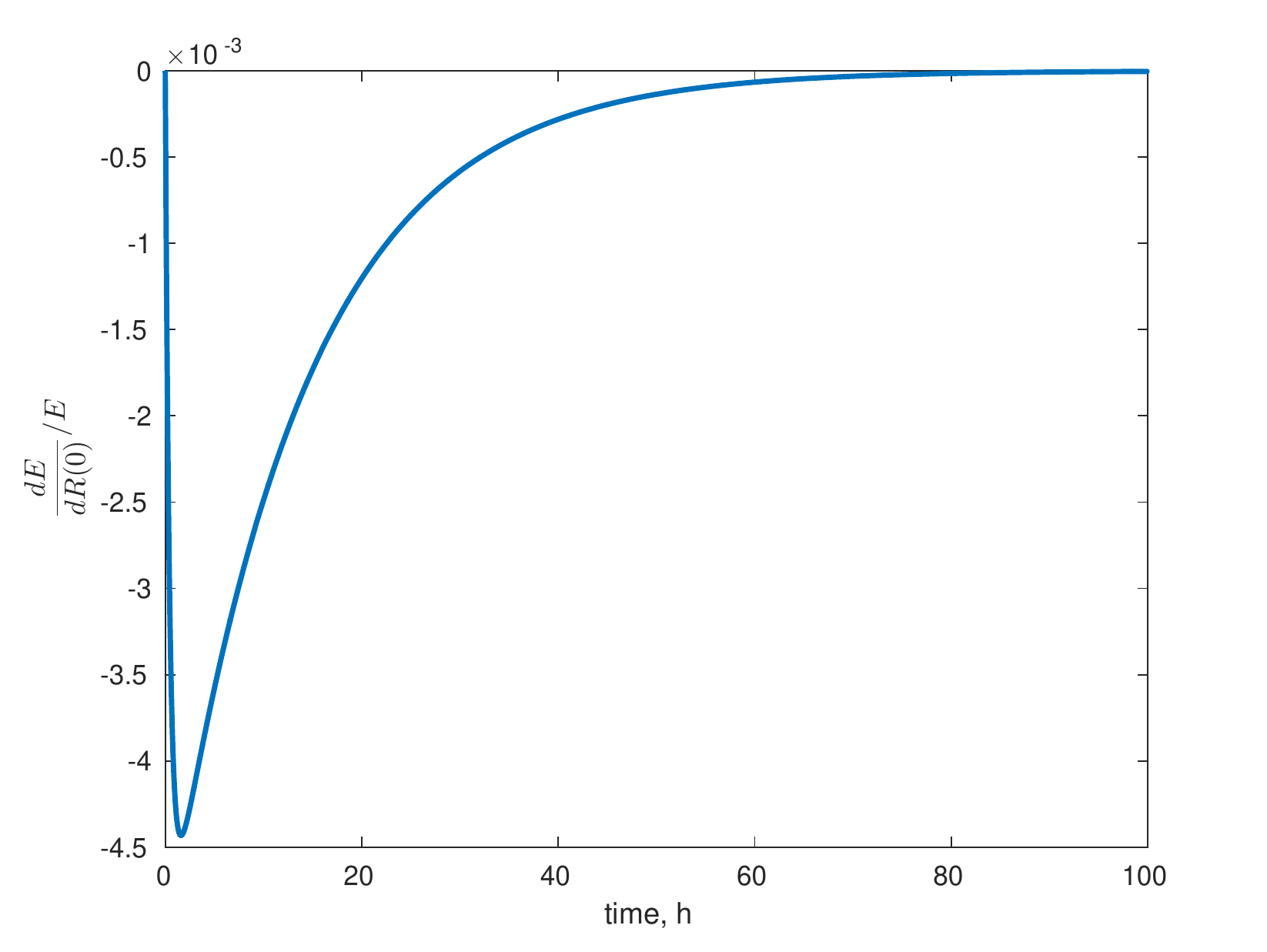}
\caption{Relative sensitivity of ATP to the initial concentration of ROS, $R(0)$.}\label{sens_E_R}  \end{figure}
%% Figure end %%

%% Figure %%
\begin{figure}[H]
\centering
\includegraphics[width=0.8\textwidth]{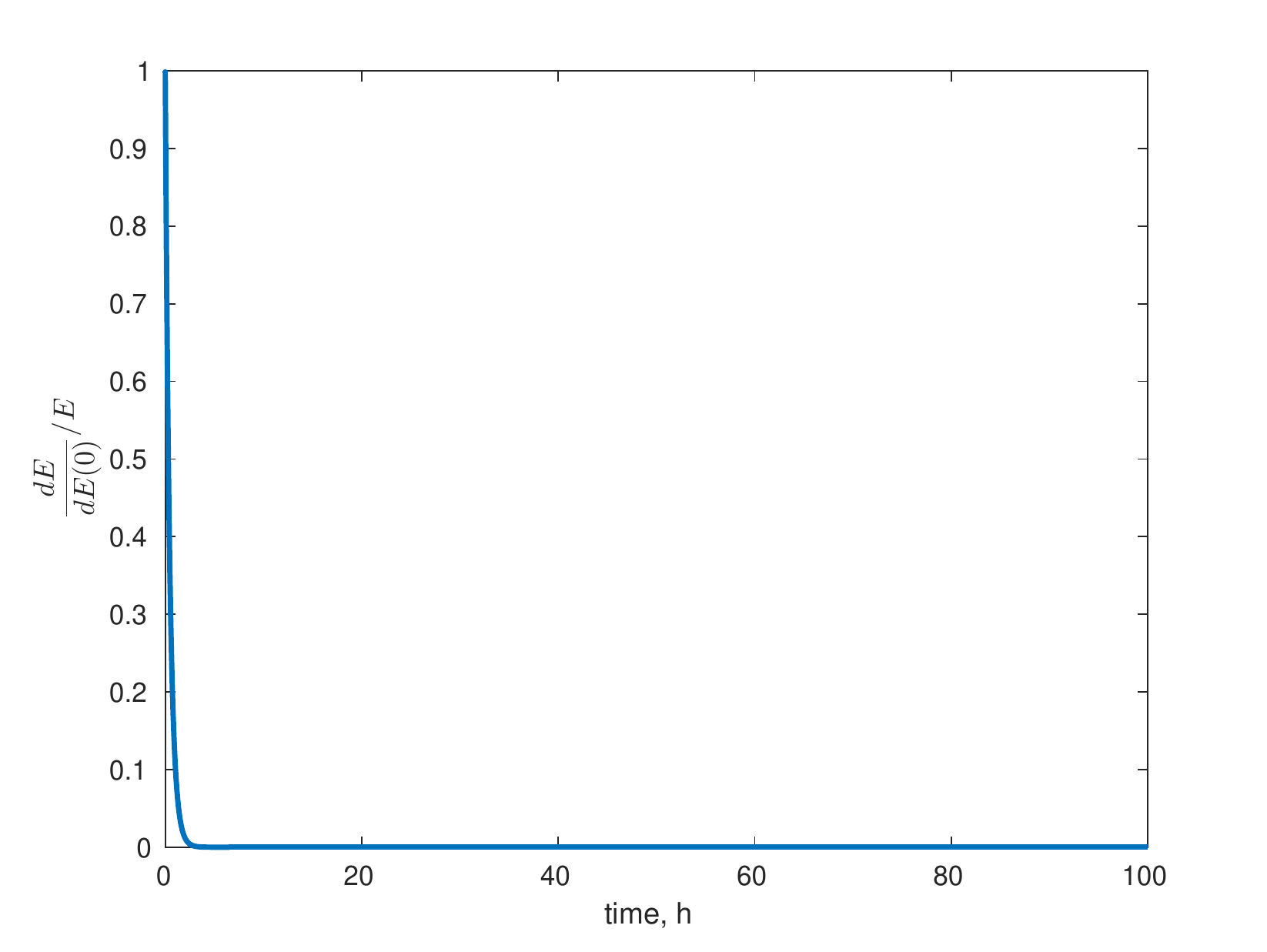}
\caption{Relative sensitivity of cells with ATP to the initial concentration of ATP, $E(0)$.}\label{sens_E_E}  \end{figure}
%% Figure end %%

\newpage

\begin{appendices}
\appendixpage
\section{Equilibrium Analysis}\label{app_eq_analysis}
The $\vec{0}$ equilibrium exists. A general equilibrium solution is:
any $M^*>0$, any $D^*>0$ (subject to the restrictions $S(R^*) = 0$ and the parameter relationships outlined in \ref{Par_rel_subs}) and
\[R^* = \frac{\alpha_M(M^* + k_DD^*)}{\delta_R},\]
\[E^* = \frac{f_E(\frac{R^*}{M^* + D^* + \epsilon})}{\delta_E},\]
%\[U^*_{1,2} = \frac{1}{2\lambda_U}(1 \pm \sqrt{1 - 4k_U\lambda_UE^*/\delta_U}).\]
%\[U^*_{1,2} = \frac{1}{2\lambda_U}(-1 \pm \sqrt{1 + 4k_U\lambda_UE^*/\delta_U}).\]
\[U^* = 0 \text{ or } U^* = \frac{1 + \lambda_UE^*}{1 + \lambda_U}.\]
\

%It is important to note that if $E^* = 0$, then $U^* = 0$ is the only equilibrium. If we assume $E^* > 0$, then we need to analyse $U_1^*$ and $U_2^*$.
\begin{comment}Let us consider $U^*_{1,2}$. If $\lambda_U > 1$, then $U^*_2$ is negative, hence can be ignored for our purposes. It is easy to show that  $U_1^*$ is stable, using the sign of $\frac{dU}{dt}$. If, however $\lambda_U < 1$, then $U_1^*$ and $U_2^*$ are both positive, $U_1^*$ is unstable and $U_2^*$ is stable. We will assume that $\lambda_U > 1$.  \end{comment}
%The only positive equilibrium is $U^* = \frac{1}{2\lambda_U}(-1 + \sqrt{1 + 4k_U\lambda_UE^*/\delta_U})$.
The $U^* = 0$ equilibrium is unstable, so if we assume that $U(0) > 0$, it will not be reached. Therefore, we only need to focus on the positive $U^*$.
As we have established, since $R^* \leq 1$, $S(R^*) = S'(R^*) = 0$. Therefore, the eigenvalues around the equilibrium are:
$e_1 = e_2 = 0$, $e_3 = -\delta_R$,$e_4 = -\delta_E$, and
\[e_5 = k_U\frac{1 + \lambda_UE^* - 2(1 + \lambda_U)U^*}{1 + \lambda_UE^*}.\]
The $U^* = 0$ equilibrium makes $e_5$ positive, as expected from the logistic nature of the $\frac{dU}{dt}$ equation in \ref{Orig_system}. Since we do not consider this equilibrium biologically viable in our case, let us consider the other equilibrium, $U^* = \frac{1 + \lambda_UE^*}{1 + \lambda_U}$. The eigenvalue $e_5 = -k_U$ in this case, so is also negative. Therefore, the equilibria are non-hyperbolic with $e_1 = e_2 = 0$ and $e_3, e_4$, and $e_5$ negative.\\
\

A simple linearization is sufficient to establish that the variables $M$ and $D$ form the center subspace of the system, while $E, U$, and $R$ form the stable subspace of the system. Because of the form of $S(R)$, the equilibrium of the system is not approached in the usual manner when considering analyses of systems of equations. Since if $R = 1$, then $\frac{dM}{dt}$ and $\frac{dD}{dt}$ are 0, which establishes $M^*$ and $D^*$. Then $R^*$ is determined by $M^*$ and $D^*$. Therefore, in order to determine the flow of the central subspace, we need to examine its behavior as $R \to 1^+$. For any $R > 1$, the system determined by the central subspace and the assumption on $R$ is:
\begin{align*}
\frac{dM}{dt} &= -k_SMs_C(R - 1)\\
\frac{dD}{dt} &= k_SMs_C(R - 1) - \delta_DDs_C(R - 1)
\end{align*}
\

%This ought to be sufficient, however it may be confusing for the reader that we would analyse this system with equilibrium ($M^*$,$D^*$).
A simple change of variables $M = M - M^*, D = D - D^*$ results in
\begin{align}
\frac{dM}{dt} &= -k_S(M + M^*)s_C(R - 1) \\
\frac{dD}{dt} &= k_S(M + M^*)s_C(R - 1) - \delta_D(D + D^*)s_C(R - 1)
\end{align}
for which $\vec{0}$ is the equilibrium (keep in mind $R \to 1^+$ is fixed). The eigenvalues of this system are $-k_Ss_C(R - 1)$ and $-\delta_Ds_C(R - 1)$, both negative. Therefore, the equilibrium ($M^*$, $D^*$) is asymptotically stable, which implies that the equilibrium of the original system is also asymptotically stable.

\end{appendices}
\end{document}